\documentclass[12pt, a4paper]{article}
\usepackage{graphicx}
\usepackage{multirow}
\usepackage{color}
\usepackage{tikz}
\usepackage{latexsym}
\usepackage{amssymb}
\usepackage{amsfonts}
\usepackage{amsmath}
\usepackage{indentfirst}
\usepackage[english]{babel}
\usepackage[utf8]{inputenc} 
\usepackage[T1]{fontenc} 
\usepackage[square,numbers]{natbib} 
\usepackage{amsmath}  
\usepackage{ellipsis}
\usepackage{listings}
\usepackage{subfig}
\usepackage{caption}
\usepackage{appendix}
\usepackage{siunitx}
\usepackage{multirow}
\usepackage{tabularx}
\usepackage{color, colortbl}

\author{Antonio Bernini, Lorella Bonaccorsi, Pietro Fanti,\\ Francesco Ranaldi, Ugo Santosuosso}
\title{Use of IT tools to search for a correlation between weather factors and onset of pulmonary thromboembolism}

\begin{document}
\date{}

\maketitle

\begin{abstract}
\textbf{Introduction}. Pulmonary embolism (PE) and deep vein thrombosis (DVT) are gathered in venous thromboembolism (VTE) and represent the third cause of cardiovascular diseases. The PE is produced by emboli, which came from venous thrombi, moving to and obstructing the arteries of the lung. PE is the most dangerous VTE resulting in significant morbidity and mortality. The PE can be fatal if undiagnosed and untreated. In PE cardiovascular risk factors associated with aging, significant immobility, surgery or trauma, active malignancy, hormone related therapies and altered haemostasis represent the main leading causes. Recent studies suggest that meteorological parameters as atmospheric pressure, temperature, and humidity could affect PE incidence but, nowadays, the relationship between these two phenomena is debated and the evidence is not completely explained. The clinical experience of the Department of Emergency Medicine at AOUC Hospital suggests the possibility that the atmospheric parameters trends are effectively related to the emergency care income of subjects diagnosed with PE. \par
\textbf{Methods}. We have collected data concerning the Emergency Medicine Unit admissions of PE patients to confirm our hypothesis. At the same time, atmospheric parameters are collected from the Laboratory for Meteorology and Environmental Modelling of Lamma Consortium of Tuscany region. We have implemented new IT models and statistic tools by using semi-hourly records of weather time high resolution data to process the dataset. We have carried out tools from econometrics, like mobile means, and we have studied anomalies through the search for peaks and possible patterns. All the study has been performed by using the high-level, general-purpose programming language, Python. We have created a framework to represent and study time series by developing a Python project and to analyze data and plot graphs. The project has been uploaded on GitHub.
Aim of the study. The aim of our study is to describe a method to evaluate the relationship between the atmospheric parameters spikes and the incomes to Emergency Unit at Careggi Hospital of subjects diagnosed with PE. To set up the study, we have retrospectively collected clinical data of PE diagnosed subjects in 2016 hospital admissions. We have set a database characterized by the demographic and pathological parameters of the subjects. In another database the meteorological data of atmospheric parameters relatively to the investigated period of time have been collected. \par
\textbf{Results}. To approach this issue we have used econometry, chronobiology (time series analysis) and pattern recognition approaches. The moving average of the time series of daily hospitalizations, on an annual window, has evidenced an increasing trend. Conversely, the annual moving average of pressure values has a decreasing trend. Our analyses highlighted a strong correlation between the moving averages of atmospheric pressure and those of the hospitalizations number although causality is still unknown. The existence of an increase in the number of hospitalizations in the days following short-to-medium periods of time characterized by a high number of half-hourly pressure changes is also detected. Average pressure changes in 4 days with half-hourly records divided by number of hospitalizations at 4th day in 2016 has an increase of 8.66\%. In the all studied period this increment was 3.18\%. The spectrograms studies obtained by the Fourier transform requires to increase the dataset. The analyzed data (especially hospitalization data) were too few to carry out this kind of analyses. \par
\textbf{Conclusions}. In conclusion, our results evidence a correlation between pulmonary embolism and meteorological parameters and in particular, atmospheric pressure (R= -0.9468, p<0,001) that is more relevant than temperature and wind speed. Further data collections will increase the investigated time and the enrolled subjects treating the information through more complex computational approaches to confirm our results. 

\end{abstract}

%\begin{keyword}
%Non-overlapping matrices, Dyck words, Catalan numbers.
%\end{keyword}

\section{Introduction}
Studying seasonality of certain diseases and an eventual influence of weather factors on their onset and mortality is a current and very debated topic, also in light of recent studies done in this sense on COVID-19's diffusion \cite{tosepu:covid}\cite{sajadi:covid}\cite{wang:covid}.\par
In particular, some recent researches seem to suggest the existence of a correlation between onset and mortality of deep vein thrombosis and/or pulmonary embolism, two diseases strongly linked together (see \ref{section:termini medici}), and weather factors: however, since the studies are recent, conclusions are often not statistically significant or in contrast to each other. Indeed some analyses highlight an increase in cases during spring months, that are characterized by a lower atmospheric pressure \cite{meral:barometric}. However, other analyses find peaks of cases in winter \cite{skajaa:venous}. Anyway, most studies agree that probably some kind of correlations exist, but more investigation is needed \cite{oztuna:meteo}.\par
The clinical experience of the Department of Emergency Medicine at AOUC Hospital seems to suggest that a connection between the number of hospitalizations for pulmonary embolism and weather factors actually exists. From this comes this research which has the purpose of using statistical tools and models to verify this correlation and, in case, to describe it. \par
Besides attempting methods already used by other researchers, we tried to tackle the problem with different and new tools, too. In particular, it was found that much of the existing literature tended to focus on studying annual and monthly means, which nevertheless have the defect of flattening values. This fact leads to a loss of information. Therefore we tried to study the problem with a higher resolution of data, using daily and semi-hourly means, the latter obtained thanks to a concession obtained by the Laboratory for Meteorology and Environmental Modelling of LaMMA Consortium of Tuscany region.\\
We have also used tools from econometrics, like mobile means, and we have studied anomalies through the search for peaks and possible patterns. \par
Everything has been realized using Python. We have created a framework to represent and study the time series. It can be used for future developments of this study or to analyze other data regarding other diseases, bot pulmonary and no, or statistic samples from different sources.

\section{Software tools and data's preliminary analysis}

\subsection{Software tools}
We have created a Python project and used it to analyse data and plot graphs which are shown in this document. The project uses \emph{pyplot} library \cite{matplotlib:pyplot} and it can be consultes and downloaded on GitHub \cite{github:weape}.\\

\subsection{Used data}
\begin{figure}
	\makebox[\textwidth][c]{\includegraphics[width=1.2\textwidth]{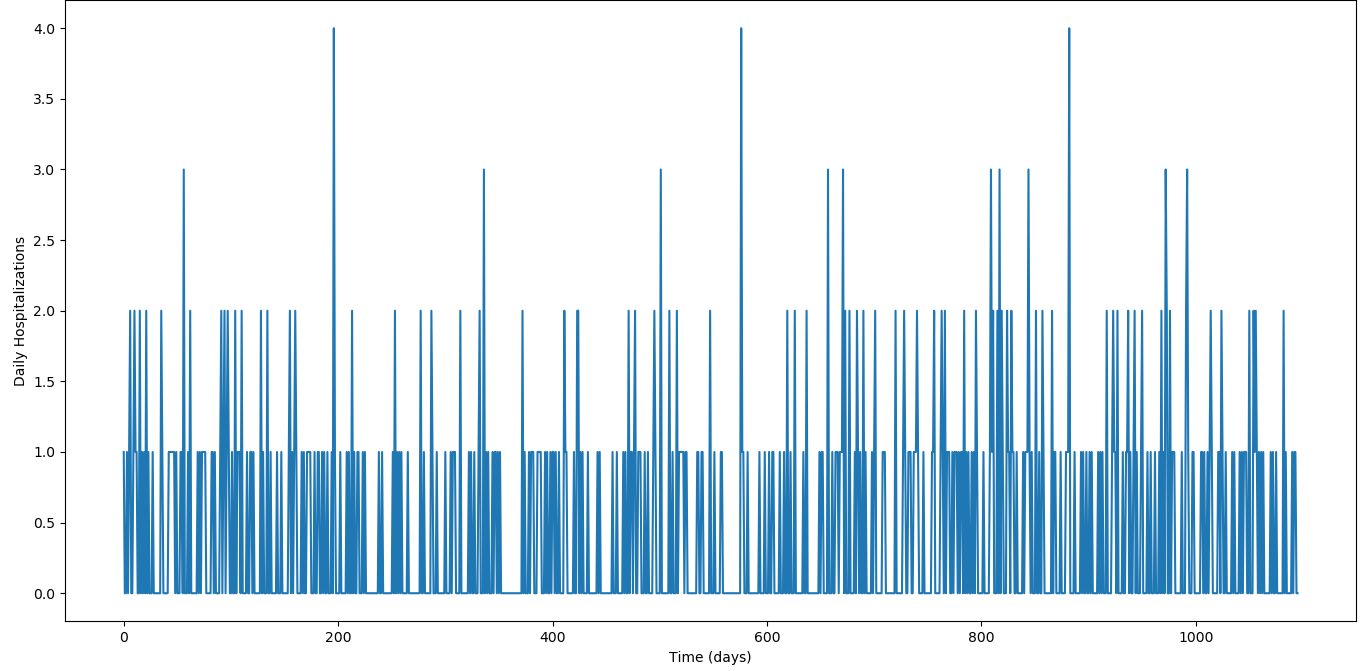}}
	\caption{Time series of daily hospitalizations (2016-2018)}
	%\captionsetup{justification=centering,margin=2cm}
	\centering
	\label{fig:hosp}	
\end{figure}

For this study we crossed data from different sources.\\
Daily hospitalizations data have been gotten by medical records of 476 cases of pulmonary embolism in the period 2016-2018, which AOUC hospital gave us. The corresponding time series is shown in Figure \ref{fig:hosp}.\par
\begin{figure}
	\makebox[\textwidth][c]{\includegraphics[width=1.2\textwidth]{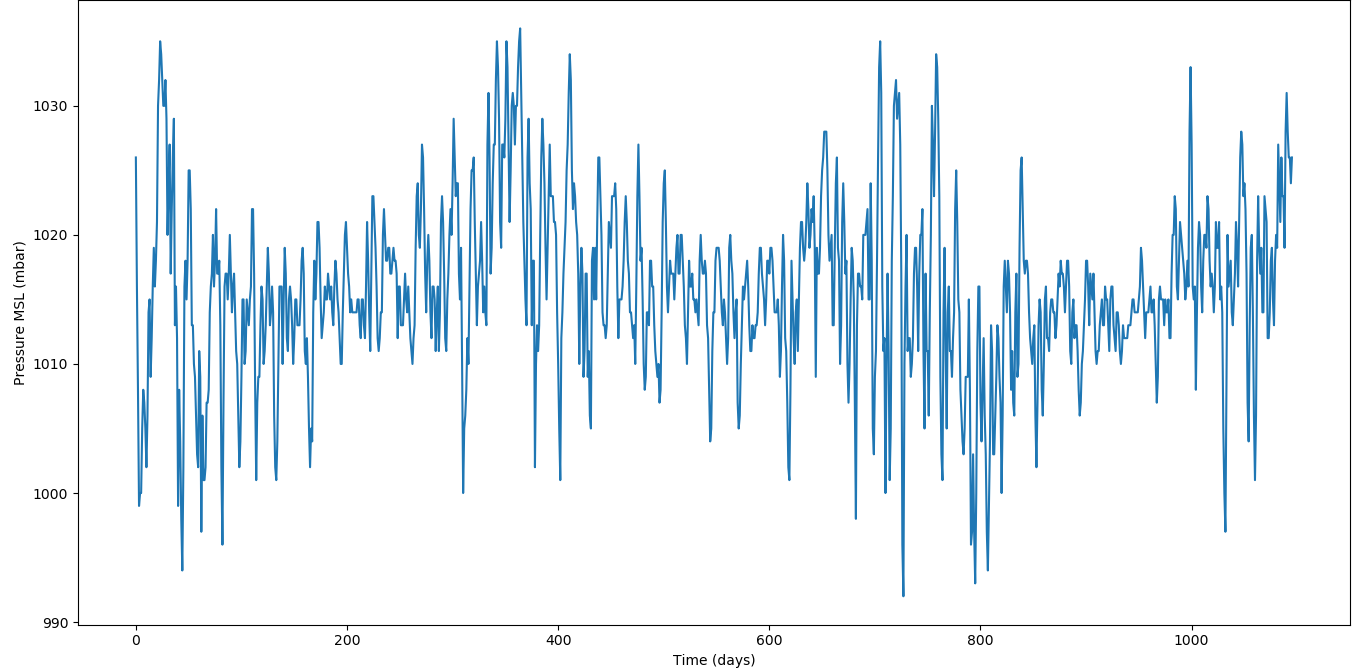}}
	\caption{Time series of daily average atmospheric pressure (2016-2018)}
%	\captionsetup{justification=centering,margin=2cm}
	\centering
	\label{fig:pres}	
\end{figure}

Daily weather data (in particular atmospheric pressure, minimum, maximum and average temperature and minimum, maximum and average wind speed) in the same period were initially obtained by \emph{meteo.it} \cite{meteo:dati}. The corresponding time series is shown in Figure \ref{fig:pres}.

	
Later, since we needed semi-hourly values rather than daily values, we have used data provided by LaMMA Consortium. These data are publicy available, only in the form of graphs, on \cite[\texttt{lamma.txt}]{github:weape}. 
For the narrow range (from August 25th 2018 to August 29th 2018) without data we have manually assigned the standard value $1000$ $mbar$.\par
Since we worked for most of the time with \emph{meteo.it}'s data, as initially we did not need semi-hourly means, in the rest of the document, when referring to weather data, always consider those taken from the latter source, unless otherwise specified.

\subsection{Preliminary Analysis}

\subsubsection{Autocorrelation}
As first approach, the autocorrelation of daily values of atmospheric pressure and, separately, of the number of hospitalizations was investigated.\par
The tool used to represent and study autocorrelation is the correlogram, also known as autocorrelogram. It is a graph constructed by examining the correlation between a time series and several dalayed series of $k$ periods, in other words the sime series $Y_t = Y_1,...,Y_T$ is correlated with his traslations of amplitude $k$. In our case, $k$ assumes values $0,1,...,10$. For every correlation, the index
$$r_k=\frac{\sum^{T}_{t=11}(Y_t - \bar{Y})(Y_{t-k}-\bar{Y})}{\sum^{T}_{t=11}(Y_t - \bar{Y})^2}$$
is calculated, whith $\bar{Y}$ representing the arithmetic mean of the values of the starting time series, not translated.
Then, the correlogram is constructed reporting in a Cartesian chart the couple of values $(k, r_k)$.\par
\begin{figure}
	\makebox[\textwidth][c]{\includegraphics[width=\textwidth]{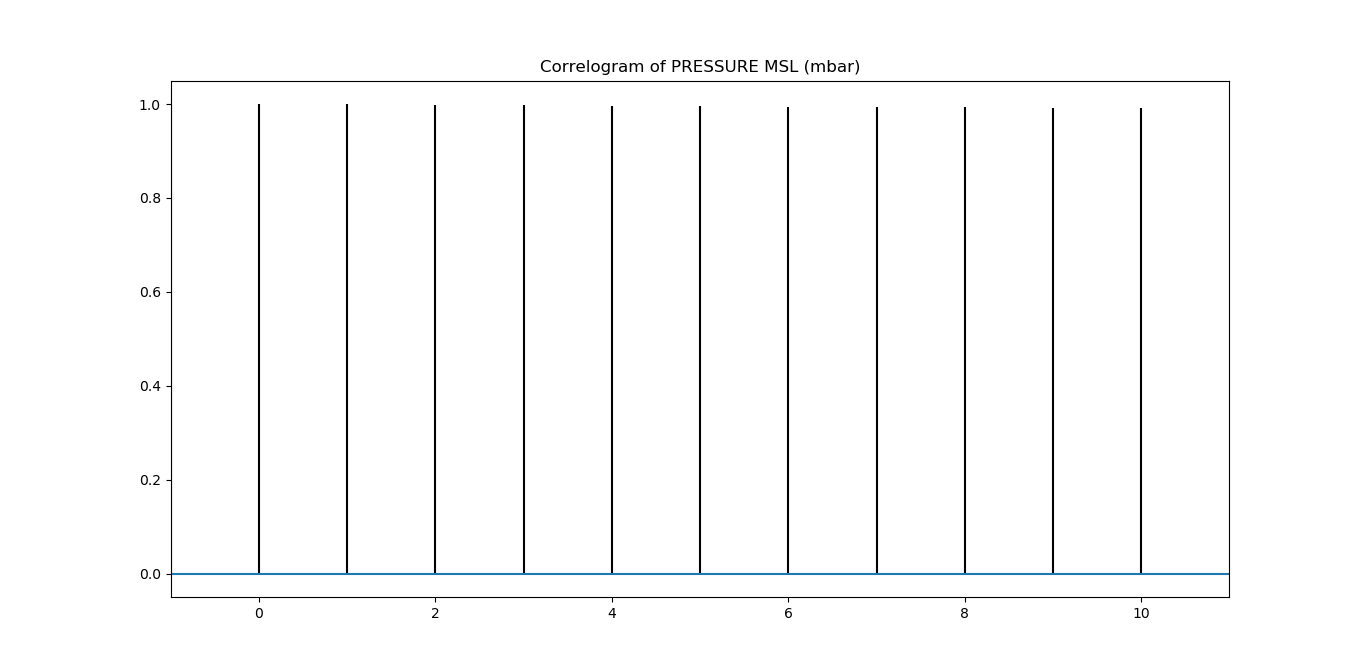}}
	\caption{Correlogram of barometric pressure values}
	%\captionsetup{justification=centering,margin=2cm}
	\centering
	\label{fig:correlogram_pres}	
\end{figure}
\begin{figure}
	\makebox[\textwidth][c]{\includegraphics[width=\textwidth]{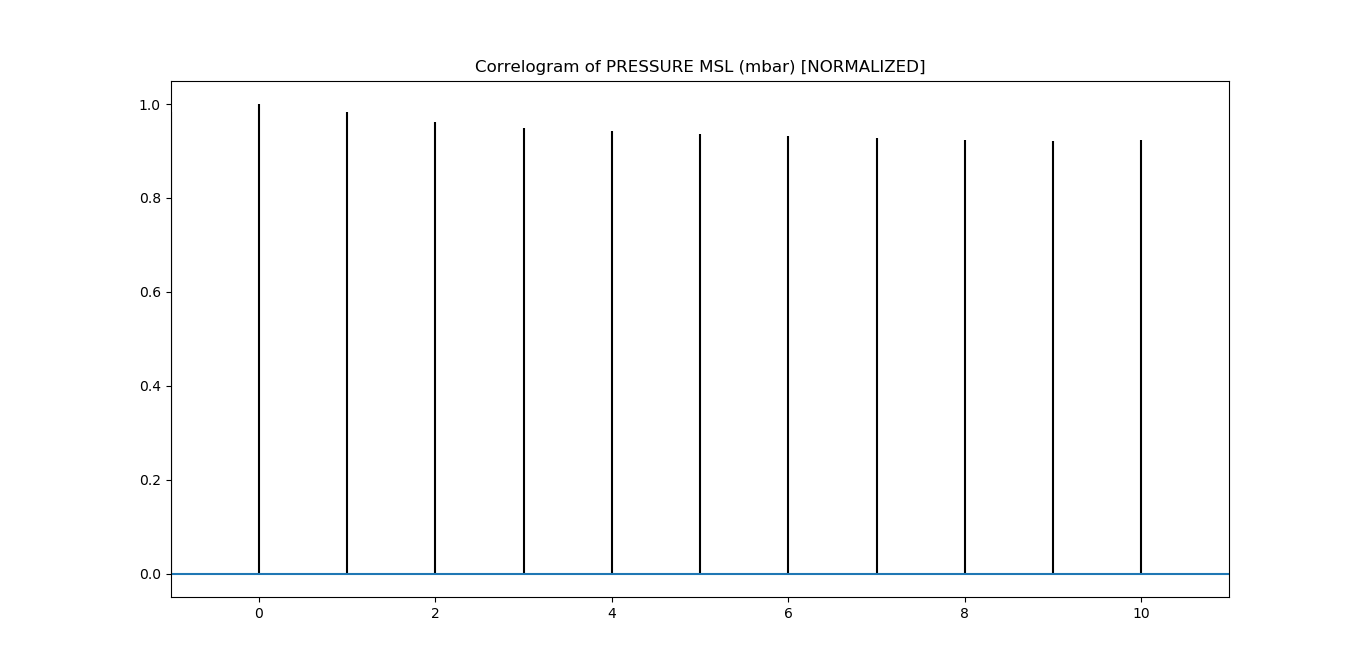}}
	\caption{Correlogram of normalized barometric pressure values}
	%\captionsetup{justification=centering,margin=2cm}
	\centering
	\label{fig:correlogram_norm}
\end{figure}
\begin{figure}
	\makebox[\textwidth][c]{\includegraphics[width=\textwidth]{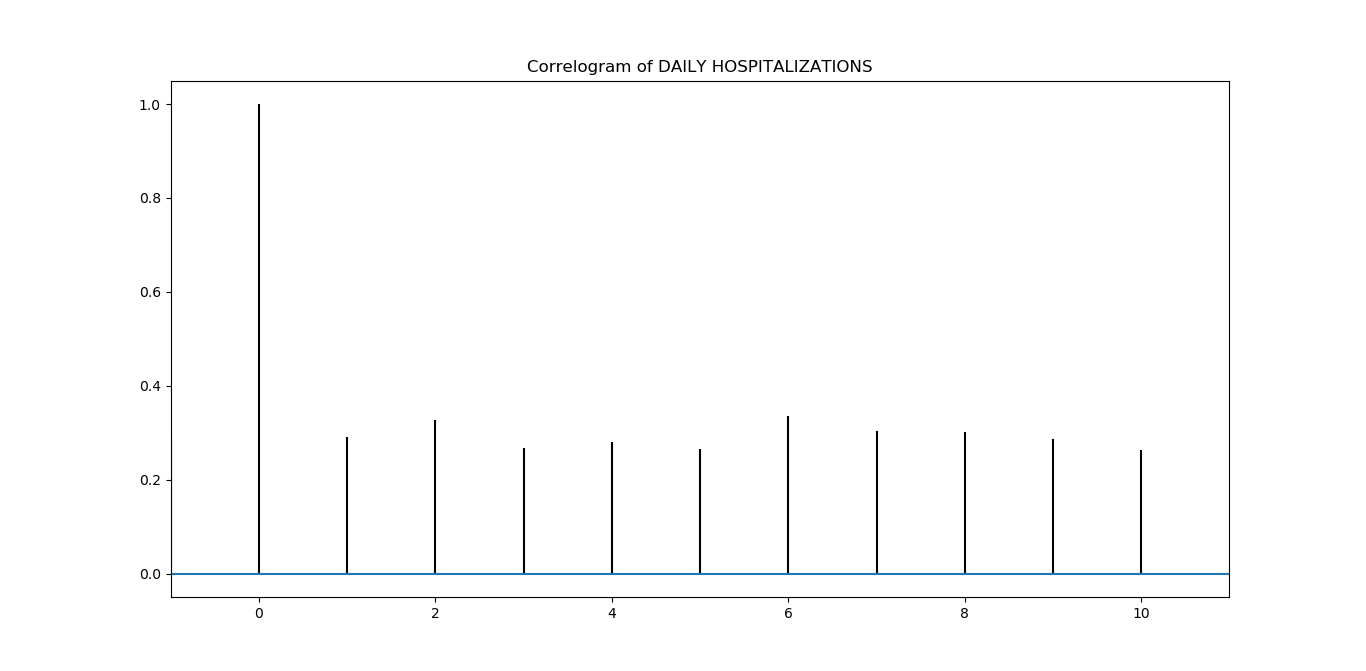}}
	\caption{Correlogram of daily hospitalizations values}
	%\captionsetup{justification=centering,margin=2cm}
	\centering
	\label{fig:correlogram_hosp}
\end{figure}

As you can see in Figure \ref{fig:correlogram_pres}, the correlogram of pressue looks flat, with a monotonically increasing trend which is highlighted making the autocorrelation of the normalized time series  (Figure \ref{fig:correlogram_norm}). For details about normalization of a time series, look at the section \ref{section:normalise}.\\
Insted the correlogram of hospitalizations, Figure \ref{fig:correlogram_hosp}, is more variegated, with a peak in $0$. This is a predictable result: as regards hospitalizations, the past has almost no influence on the present and the accidental component prevails. In the pressure values, on the contrary, the trend component prevails, with the present being strongly influenced by the past, in particular by the nearest one, with decreasing influence as we move backwards, as evidenced by the decreasing correlogram \cite{fonzo:serie}.\par
For the algorithm used, see \cite[\texttt{auto\_correlation.py}]{github:weape}.


\subsubsection{Seasonal adjustment}
\label{section:destagionalizzazione}
\begin{figure}
	\makebox[\textwidth][c]{\includegraphics[width=\textwidth]{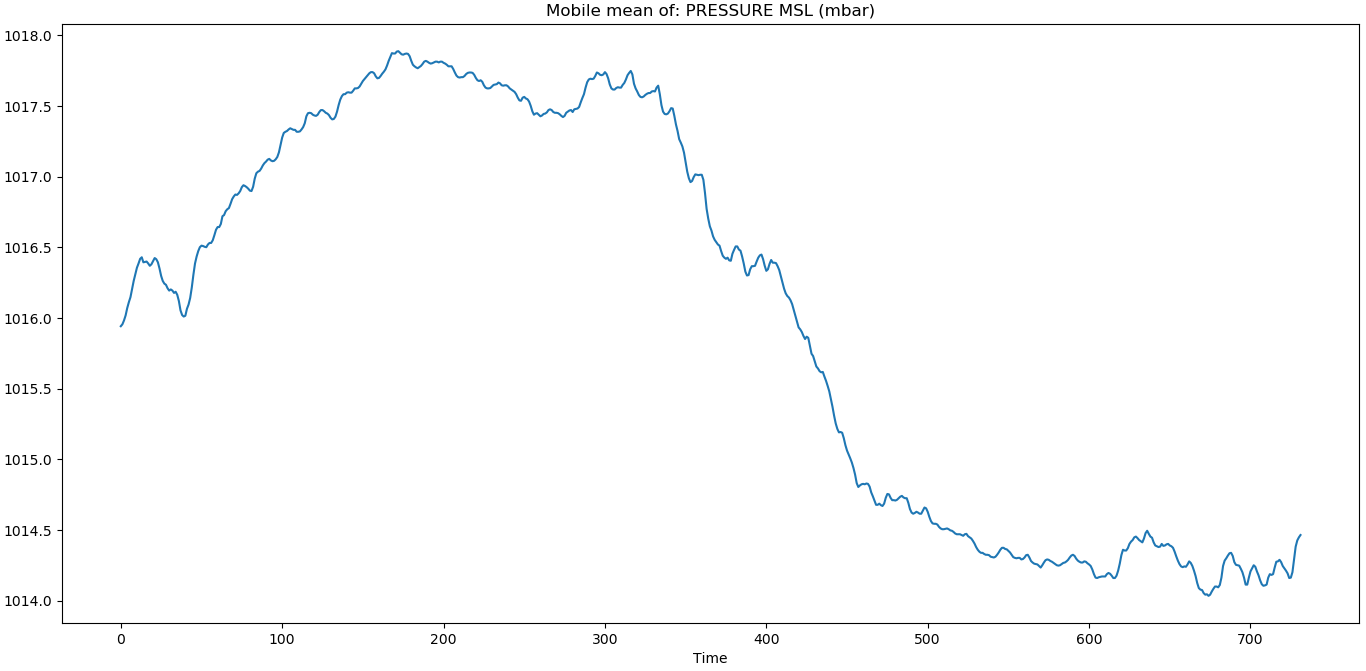}}
	\caption{Annual moving average of pressure values}
	%\captionsetup{justification=centering,margin=2cm}
	\centering
	\label{fig:media_pres}
\end{figure}

Since in the graph in the Figure \ref{fig:pres} there are peaks of annual seasonality, in an attempt to seasonally adjust the series, its moving average was realized on a 365-day window. The resulting historical series, shown in the Figure \ref{fig:media_pres}, has a decreasing trend. Using the data provided by LaMMA, a similar graph is obtained, so this historical series can be considered completely reliable. The trend observed is not surprisingly: barometric variations of this amplitude are in fact normal in the time spans considered.\par

\begin{figure}
	\makebox[\textwidth][c]{\includegraphics[width=\textwidth]{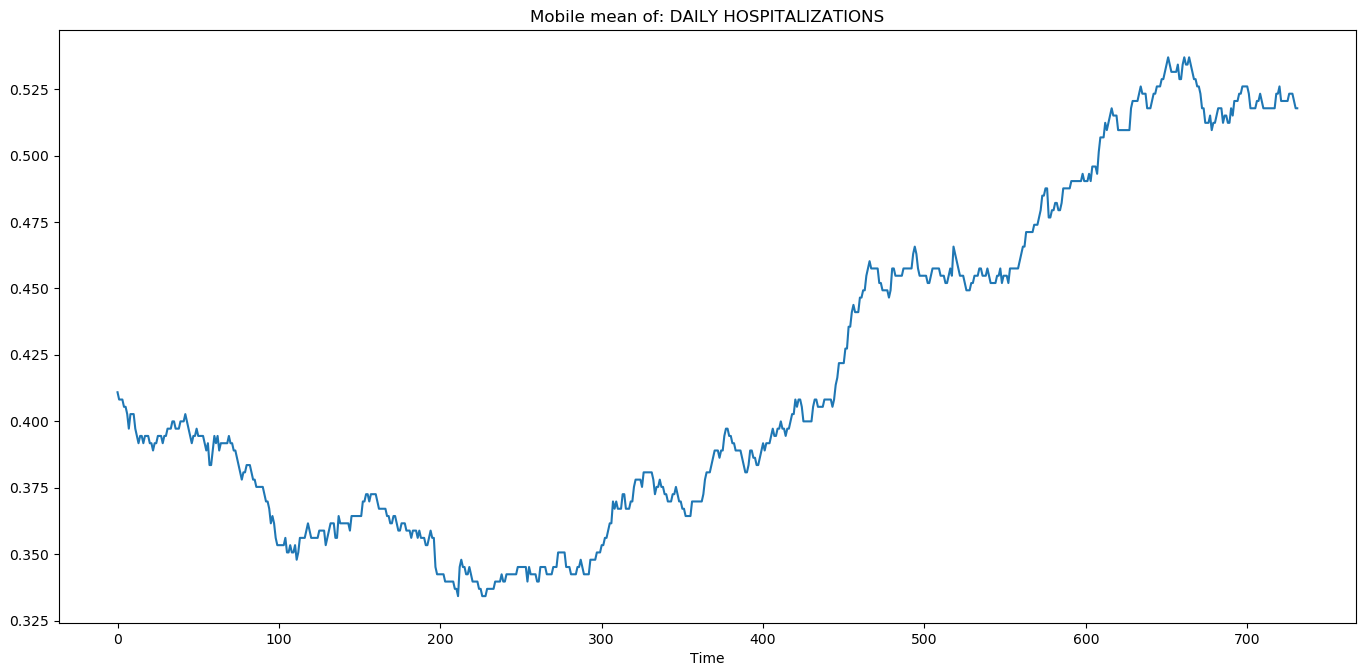}}
	\caption{Annual moving average of the number of hospitalizations}
	%\captionsetup{justification=centering,margin=2cm}
	\centering
	\label{fig:media_hosp}
\end{figure}

By making the moving average of the time series of daily hospitalizations, always on an annual window, we obtain a graph (Figure \ref{fig:media_hosp}) with an increasing trend, which cannot be ignored in the analysis of the results obtained later. Some hypotheses about the cause of this growing trend include:
\begin{itemize}
	\item improvement of diagnostic tools;
	\item increased awareness of the pathology, with consequent increase in the frequency of diagnosis;
	\item increased precription of drugs with pulmonary embolism as a side effect;
	\item increase of triggering factors such as atmospheric phenomena and/or polluting agents.
\end{itemize}
Note that only the last of the hypotheses formulated above can lead to finding a correlation between the data in our possession. If, on the other hand, the cause was to be found in one of the other hypothesised phenomena (or were in any case of another nature), the growing trend found would make the data less "clear" for the purposes of the study, constituting an obstacle in data analysis.

\section{Search for a correlation between atmospheric pressure and number of hospitalizations}
%\chaptermark{Correlation between pressure and number of
% hospitalizations}

\subsection{Correlation coefficients used}
The most immediate tool for finding a correlation is the scatter plot, however, if used alone, it can be conditioned by subjective interpretations. For this reason, the use of this tool has been accompanied by two of the most widely used correlation coefficients.

\subsubsection{Pearson correlation coefficient}
The Pearson correlation coefficient, or Pearson's r, measures the degree of linear correlation between two statistical variables. Like other correlation coefficients it assumes values belonging to the range $[- 1, 1]$, where $0$ implies the absence of a correlation, while $1$ and $-1$ indicate a linear correlation, respectively, positive and negative. Intermediate values indicate a linear correlation (positive or negative depending on the sign) more or less strong depending on how much they deviate from $0$.\\
The coefficient is calculated by the formula

	$$r = \frac{\sum(x - m_x)(y - m_y)}{\sqrt{\sum(x-m_x)^2\sum(y-m_y)^2}}$$

where $x$ and $y$ represent respectively the values assumed by the two random variables $X$ and $Y$ whose correlation we are studying, while $m_x$ and $m_y$ represent the averages.\par
The $p$ value associated with the calculation of the Pearson coefficient indicates the probability that a randomly generated system of variables has a degree of correlation at least equal to that of the system examined. Note however that the calculation of $p$ is completely correct only under the assumption that all the variables examined have normal distribution, which is not always ensured in the analyzes made below, which is why the values of $p$ reported could be subject to errors \cite{scipy:fft}.

\subsubsection{Spearman's rank correlation coefficient}
The Spearman's rank correlation coefficient, or Spearman's coefficient, indicates the degree of monotonic correlation between two random variables \cite{spearman:proof}. Like the Pearson coefficient, it assumes values belonging to the range $[- 1, 1]$ where $0$ indicates the absence of a monotonic correlation, while $-1$ and $1$, on the contrary, indicate an exact monotonic correlation , negative in the first case, positive in the second. It is defined as a particular case of the Pearson coefficient in which the values are converted into ranks before moving on to the calculation of the coefficient \cite{myers:research}.\par
Compared to the Pearson coefficient, it has the advantage of also detecting non-linear correlations, albeit always monotonic, and of not needing the normality of the distribution of the variables for the calculation of the associated $p$ value, which is however sufficiently reliable only for sets of data of at least 500 values \cite{scipy:fft}, hypothesis that however, unlike what happened for the Pearson coefficient, is always satisfied in the present study.

\subsection{Attempts made}
\label{section:correlazione}
\begin{figure}
	\makebox[\textwidth][c]{\includegraphics[width=\textwidth]{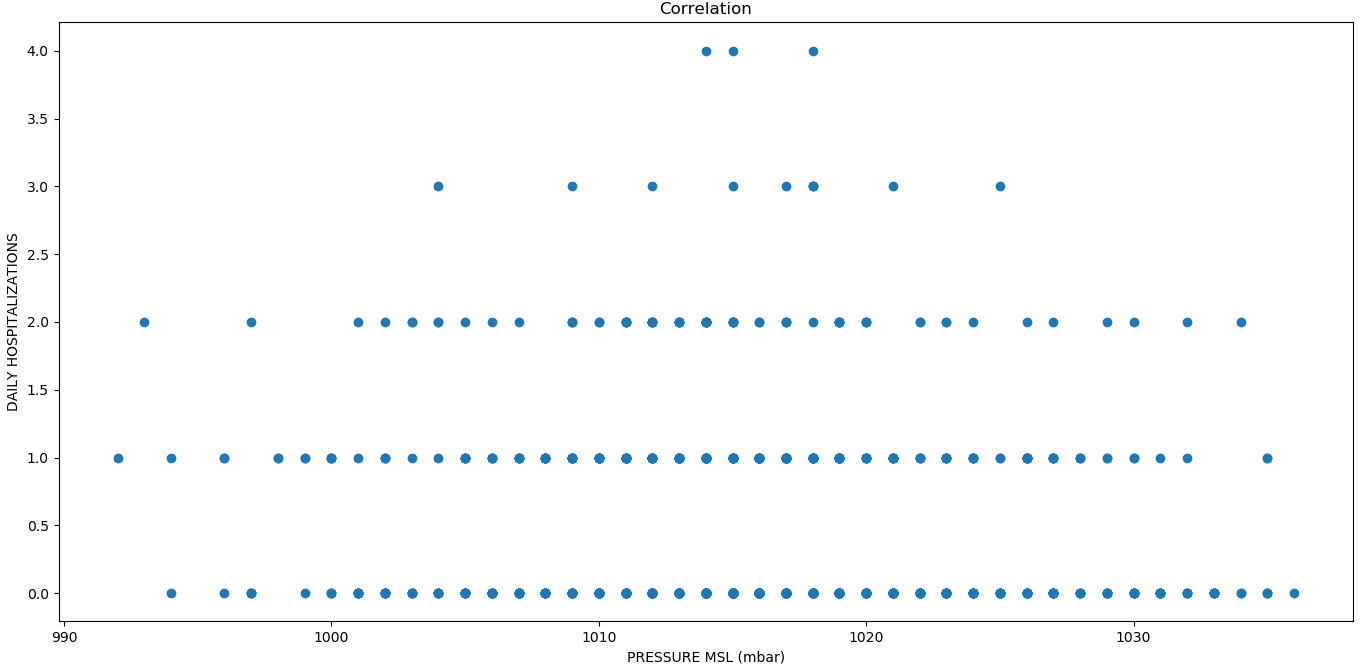}}
	\caption{Scatter plot between pressure and hospitalizations}
%	\captionsetup{justification=centering,margin=2cm}
	\centering
	\label{fig:corr_pres}
\end{figure}

A first approach in the search for a correlation between the time series of hospitalizations and that of atmospheric pressure was to generate a scatter plot (Figure \ref{fig:corr_pres}), which however did not highlight anything in particular, but, at first sight, it shows a fairly random relationship between the data.\par
\begin{figure}
	\makebox[\textwidth][c]{\includegraphics[width=\textwidth]{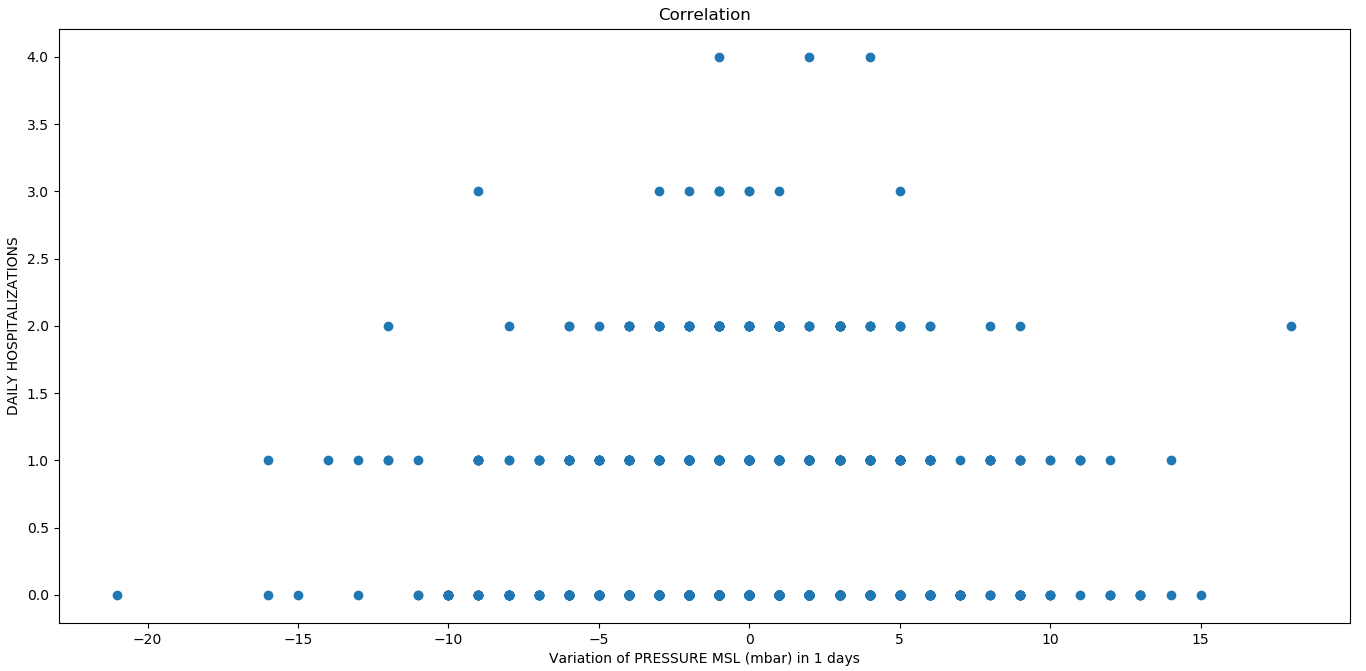}}
	\caption{Scatter plot between pressure variation and hospitalizations}
	%\captionsetup{justification=centering,margin=2cm}
	\centering
	\label{fig:corr_pres_var}
\end{figure}

Something that, at least apparently, seems more relevant is what is shown in Figure \ref{fig:corr_pres_var} where a scatter plot between the time series of hospitalizations and the time series of the variation of pressure values since one day to another has been generated. The resulting graph appears to deviate slightly more from a completely random distribution, with the upper left part of the figure completely empty, however, trying to calculate the Pearson and Spearman coefficients, no relevant results were produced. For the algorithm used, see \cite[\texttt{corr\_pressure.py}]{github:weape}.\\
Note that in both Figures \ref{fig:corr_pres} and \ref {fig:corr_pres_var} many of the points actually represent multiple occurrences, not visible because they overlap.\par
\begin{figure}
	\makebox[\textwidth][c]{\includegraphics[width=\textwidth]{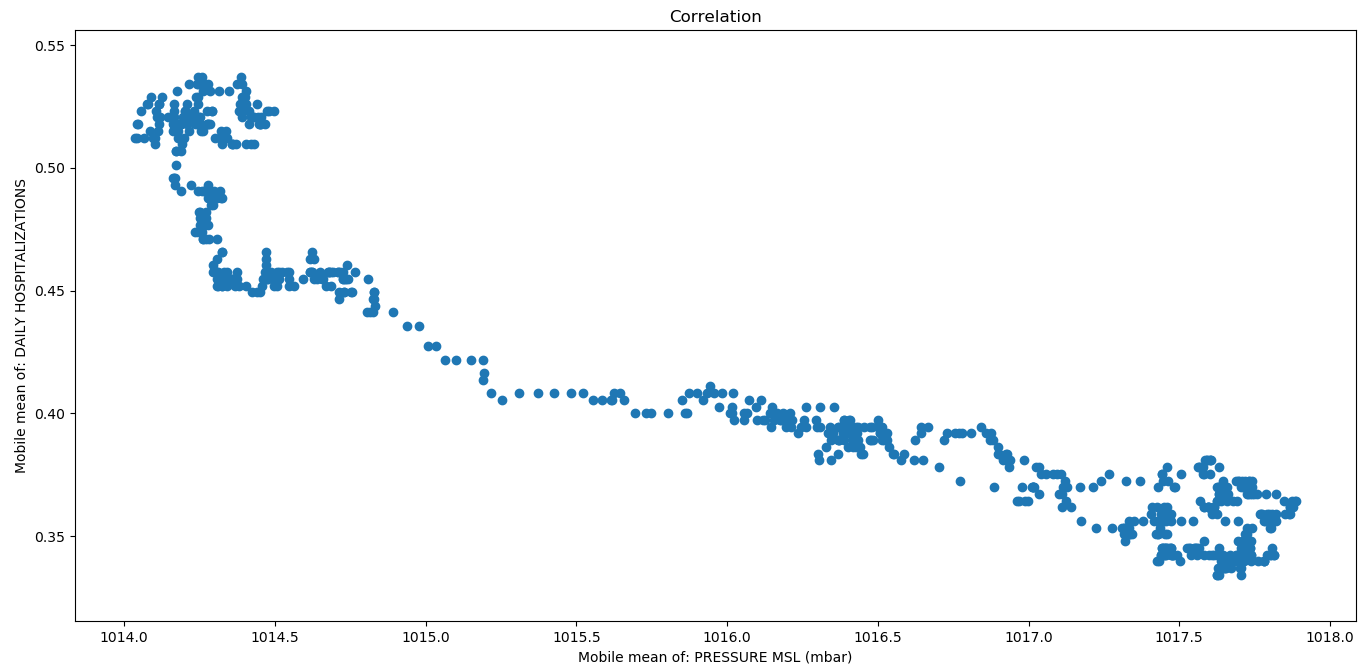}}
	\caption{Scatter plot between annual moving averages of pressure and hospitalizations}
	%\captionsetup{justification=centering,margin=2cm}
	\centering
	\label{fig:mob_mean_corr}
\end{figure}

 Correlating the seasonally adjusted time series mentioned in the section \ref{section:destagionalizzazione}, the result obtained is completely different. The dispersion graph in Figure \ref{fig:mob_mean_corr} shows a negative linear correlation, confirmed by the Pearson ($-0.9468, p = 0.0$) and Spearman ($-0.9324, p = 0.0$) coefficients, both close to $-1$. This result is in line with studies in the literature which show a higher incidence of pulmonary embolism in the months characterized by low atmospheric pressure \cite{meral:barometric}.\par
\begin{figure}
	\centering
	\subfloat[][\emph{7 days}]
		{\includegraphics[width=.45\textwidth]{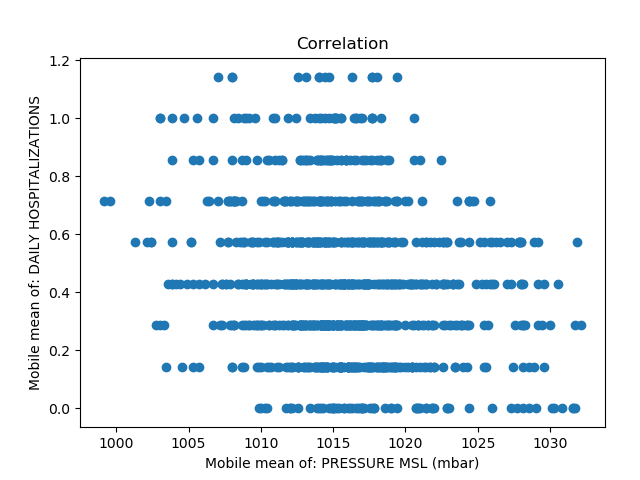}} \quad
	\subfloat[][\emph{30 days}]
		{\includegraphics[width=.45\textwidth]{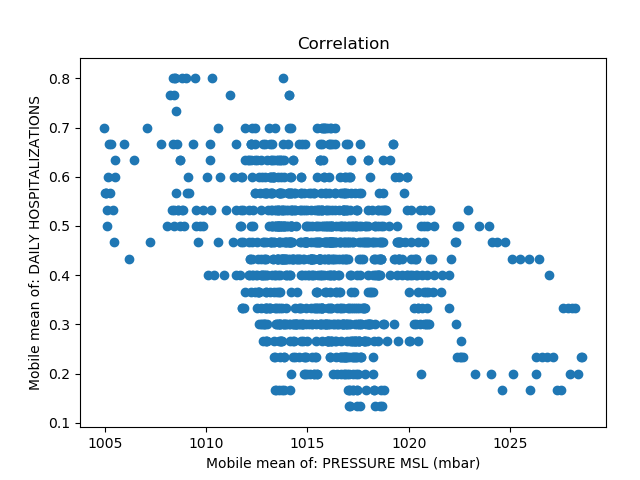}} \\
	\subfloat[][\emph{120 days}]
		{\includegraphics[width=.45\textwidth]{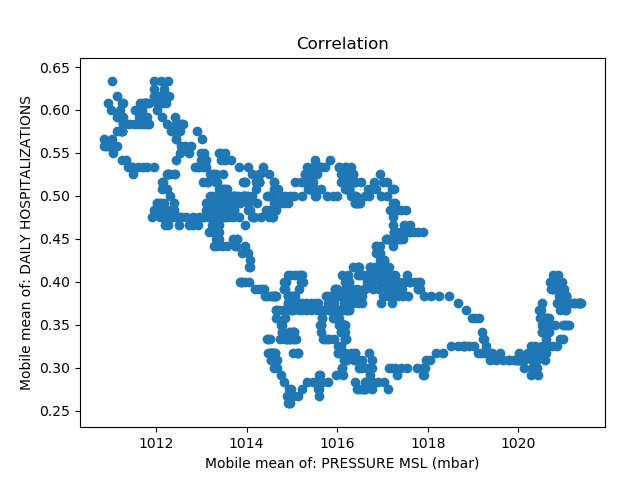}} \quad
	\subfloat[][\emph{365 days}]
		{\includegraphics[width=.45\textwidth]{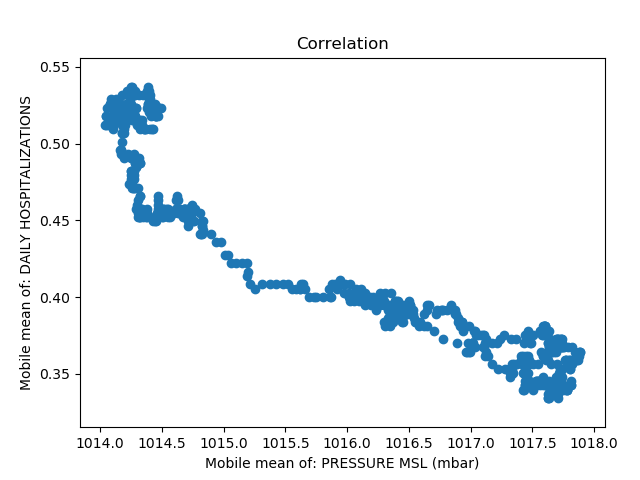}}
	\caption{Scatter plots between atmospheric pressure and hospitalizations with moving averages over intervals of increasing amplitude}
	\label{fig:medie_mobili}
\end{figure}

 In general, there is an increase in the negative correlation as the width of the moving average range increases, this is highlighted by the graphs in Figure \ref{fig:medie_mobili} (notice that the graph (d) is the same as shown in Figure \ref{fig:mob_mean_corr}) and its correlation ceofficientes in Table \ref{tab:medie}. For the algorithm used, see \cite[\texttt{corr\_mobile\_means.py}]{github:weape}. \par Although a correlation regarding seasonally adjusted values is undeniable, causality is not necessarily so. In fact, we have already discussed in the section \ref{section:destagionalizzazione} about the possible causes of the increase in hospitalizations, and several have been hypothesized that have nothing to do with atmospheric pressure. Investigations on this causality and possible explanations will be the subject of future research in the context of a collaboration with biomedical and experimental doctors.

\begin{table}
\begin{tabularx}{\textwidth}{X X X X}
%\toprule
\hline
Window lenght & Coefficient & Value & p\\
\hline
%\midrule
\multirow{2}*{7 days} & Pearson & $-0.2197$ & $<0.001$ \\
& Spearman & $-0.2052$ & $<0.001$ \\
%\midrule
\hline
\multirow{2}*{30 days} & Pearson & $-0.3676$ & $<0.001$ \\
& Spearman & $-0.3222$ & $<0.001$ \\
%\midrule
\hline
\multirow{2}*{120 days} & Pearson & $-0.6311$ & $<0.001$ \\
& Spearman & $-0.6018$ & $<0.001$ \\
%\midrule
\hline
\multirow{2}*{365 days} & Pearson & $-0.9468$ & $0.0$ \\
& Spearman & $-0.9324$ & $0.0$ \\
\hline
%\bottomrule
\end{tabularx}
\caption{Correlation coefficients depending on the length of the moving average windows considered}
\label{tab:medie}
\end{table}

\section{Peaks analysis}
Given the nature of the biological phenomenon that causes pulmonary embolisms and
the possible correlation with environmental factors (see Appendix B), it is reasonable to consider the analysis of the peaks of the historical series of such data, and their patterns, as a valid tool for look for some kind of correlation between data. It was initially thought that a sudden  pressure jump could cause the detachment of the thrombus with the consequent appearing of the embolus (see B.1). \ref{section:termini medici}).

\subsection{Peak definition}
Let $A = a_0, a_1, ..., a_n$ a time series. Let $w \in \mathbb{N}$ with $w > 0$ and  $f \in \mathbb{R}$ with $f > 0$. We indicate by $mean:  \mathbb{R}^w \to  \mathbb{R}$  and  $std:  \mathbb{R}^w \to  \mathbb{R}$ the arithmetic mean and the mean squared error functions, respectively. We define the set $\{P^+\}$ of positive peaks as:
	$$P^+ = \{a_i \in A | a_i > mean(a_{i-w+1}, ..., a_i) + f  std(a_{i-w+1}, ..., a_i)\}$$
Similarly, the set $\{P^-\}$ of negative peaks is:

	$$P^- = \{a_i \in A | a_i < mean(a_{i-w+1}, ..., a_i) - f  std(a_{i-w+1}, ..., a_i)\}$$
	
After testing different combinations of values of $w$ and $f$ it was considered optimal the choice $w=7$ and $f=1$,regarding
the quantity of detected peaks and their relevance for the purposes 
of our research.

\subsection{Patterns analysis}
Once the time series has been transformed into a peak time series,
%by simply setting $1$ and $-1$ as the values ​​that are, respectively,
by simply setting $1$ and $-1$ as the values that are, respectively,
positive or negative peaks, and equal to $0$ all other values, it is possible to
obtain a historical series of patterns. This is achieved by analyzing
all the consecutive n-tuples (with n equal to the length of the considered pattern) of peaks series (which is a sequence of $-1$, $0$ and $1$ values), and posing $1$ for each occurrence of the pattern, $0$ otherwise.

\begin{figure}
	\makebox[\textwidth][c]{\includegraphics[width=\textwidth]{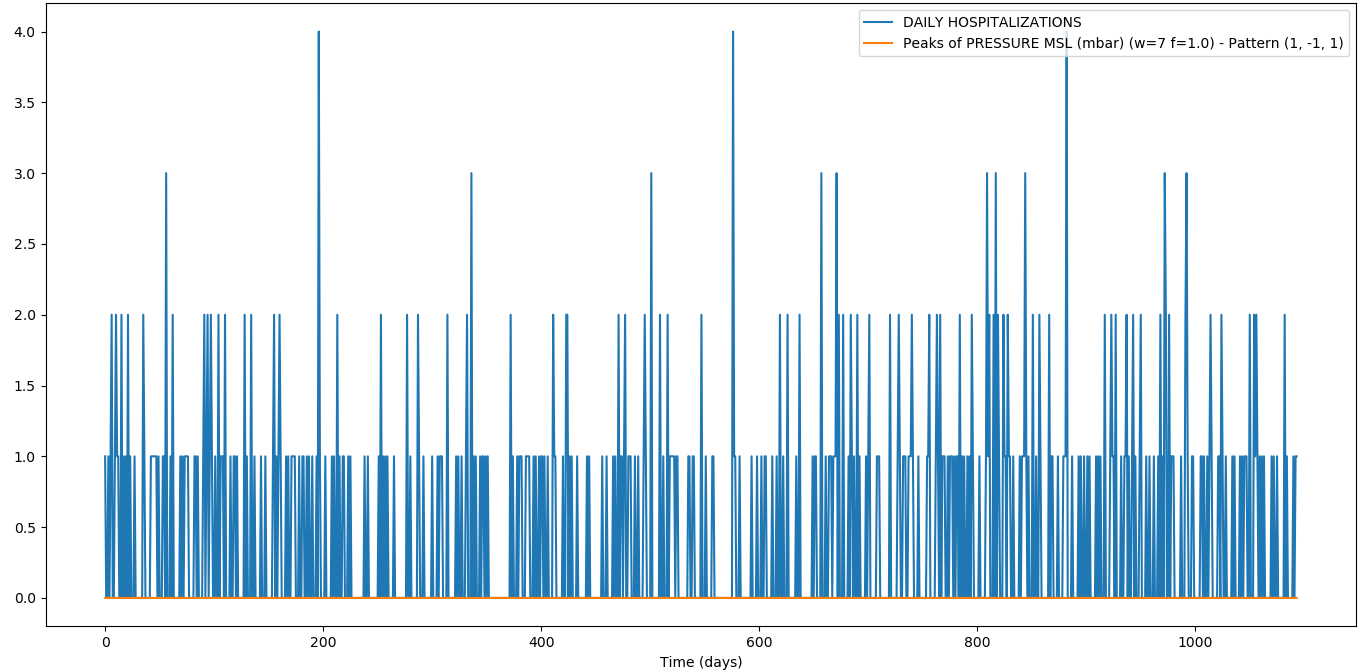}}
	\caption{Occurences of the  (1,-1,1) pattern with respect to the  
	hospitalizations}
%	\captionsetup{justification=centering,margin=2cm}
	\centering
	\label{fig:pattern_+-+}
\end{figure}

\begin{figure}
	\makebox[\textwidth][c]{\includegraphics[width=\textwidth]{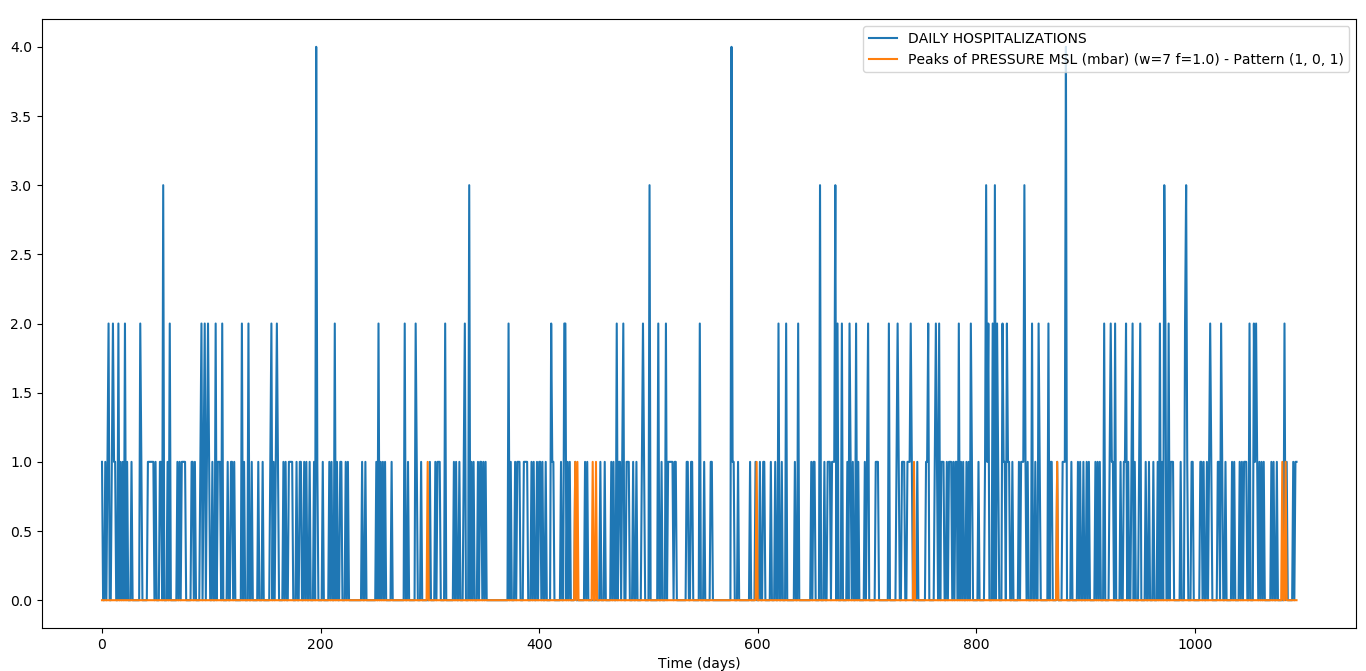}}
	\caption{Occurences of the  (1,0,1) pattern with respect to the  
	hospitalizations}
	%\captionsetup{justification=centering,margin=2cm}
	\centering
	\label{fig:pattern_+0+}
\end{figure}

\begin{figure}
	\makebox[\textwidth][c]{\includegraphics[width=\textwidth]{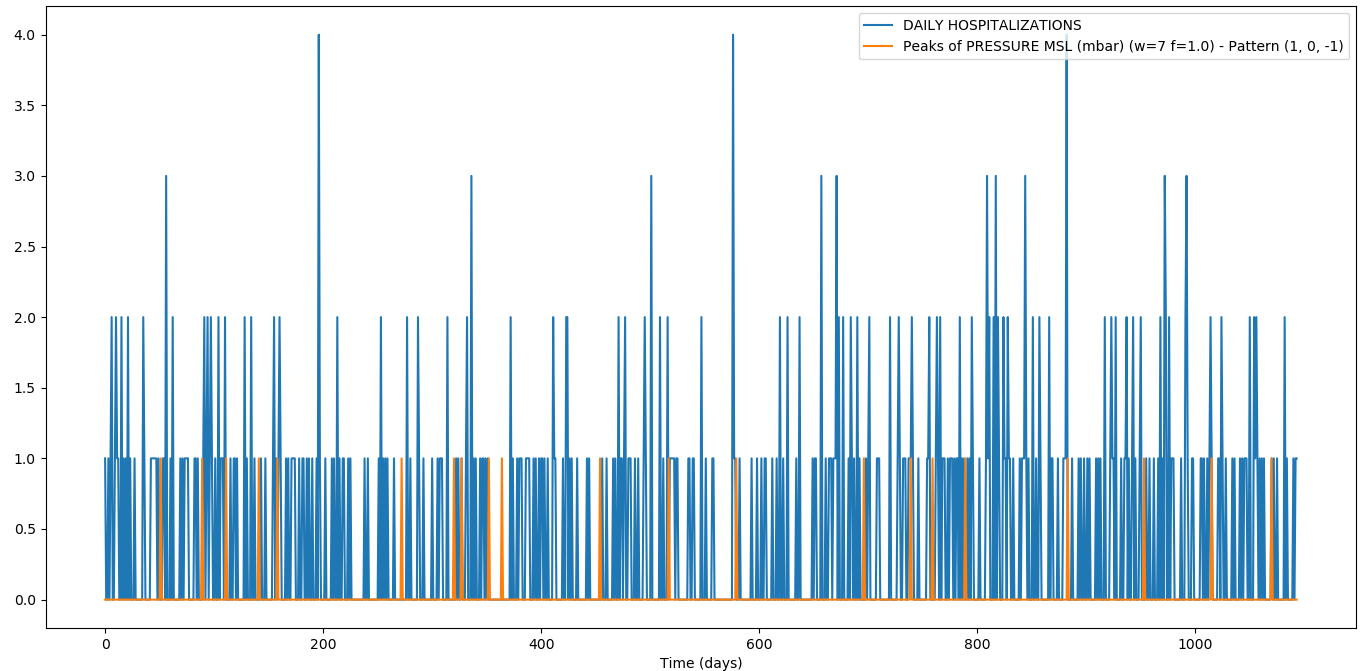}}
	\caption{Occurences of the  (1,0,-1) pattern with respect to the  
	hospitalizations}
	%\captionsetup{justification=centering,margin=2cm}
	\centering
	\label{fig:pattern_+0-}
\end{figure}

Note that, since w=7 and n=3, the occurrence of a pattern involves an interval of 9 consecutive days which are the day the pattern is found and the 8 previous days.
Figures \ref{fig:pattern_+-+}, \ref{fig:pattern_+0+} e \ref{fig:pattern_+0-}, obtained by performing \cite[\texttt{pattern\_hospitalizations.py}]{github:weape}, show a comparison between the incidence
of some patterns of the atmospheric pressure peaks
and the number of daily hospitalizations. The chosen patterns are $(1,-1,1)$, $(1,0,1)$, and $(1,0,-1)$ since they reveal the greatest pressure changes
in restricted periods. Notice how the $(1,-1,1)$ pattern never 
occurs, so that we deduce that in the geographical area of study
the atmospheric pressure does not 
change too quickly or at least this did not happen in the period of 
interest.

\subsubsection{Example}
Consider the following values:
$$
\{900,900,900,900,900,900,1020,900,1000,900\}
$$

With $w=7$, the corresponding sequence of the moving average is

$$
\{917.14, 917.14, 931.43, 931.43\}
$$

We then calculate the absolute value of the difference between the last value of
each interval and its moving average:

$$
\{|1020 - 917.14|, |900 - 917.14|, |1000 - 931.43|, |900 - 931.43|\}=\{102.86, 17.14, 68.57, 31.43\}
$$

Considering the sequence of mean squared errors
$$
\{41.99; 41.99; 49.97; 49.97\}
$$
we obtain the peaks time series:

$$
\{1,0,1,0\}.
$$

If we are looking for the pattern $(1,0,1)$ we obtain the following pattern time series:
$$
\{1,0\}.
$$

Note that the same results would have been obtained even if the
values 1000 and 1020 had been reversed.

\subsubsection{Final considerations}
At least with the data in our possession, it is difficult to describe
a relationship between the occurrence of a certain pattern (in the graphs
represented by vertical segments in orange) and the number of hospitalizations
in the next days. For this reason,
in order to better study the effect
pressure surges on the number of hospitalizations we have used also other tools (described in Section \ref{section:sbalzi}), which needed semi-hourly pressure data.

\section{Automatic search of a linear correlation or a correlation described by a monotone function}
%\chaptermark{Ricerca automatizzata di una correlazione %monotona}

File \cite[\texttt{correlation\_finder.py}]{github:weape} implements an algorithm for an automatic search of correlation having the Spearman or Pearson cofficient as large as possible.
The algorithm takes into account the following times series: atmospheric pressure, daily minimal, mamximum and average temperature, average and amximum wind speed.
Each of these series is compared with the historical series of hospitalizations,
both directly and after the applications of some transformations discussed in the previous chapters. More precisely:

\begin{itemize}
	\item deduce the times series of variations on time interval from 1 to 7 (see A.3.2) \ref{section:variationseries}; 
	\item deduce the time series of peaks and look for a correlation applaying different translations on the considered time series and the time series of hospitalizations;
	\item the historical series of the patterns $(1; -1; 1)$, $(1; 0; 1)$ and $(1; 0; -1)$  are derived from the historical series of the peaks and then only the
historical series of patterns having at least one occurrence of the considered pattern are taken into account.
\end{itemize}
Almost all parameters are easily modifiable, in particular:
\begin{itemize}
	\item meteorological data to be considered;
	\item length of the intervals to generate the time series of the
variations;
%	\item $w$ and $f$ values ​​for the search of peaks;
	\item $w$ and $f$ values for the search of peaks;
	\item length of the maximum translation between the time series of the  hospitalizations and peak time series;
	\item pattern of peaks to search.
\end{itemize}
\par
With the configuration we used to run our algorithm, the correlations
examined amounted to 138. Then, the algorithm returns a graph
of comparison between the time series, the dispersion graph and the indices of
correlation of the that series which, after the correlation with daily hospitalizations, has the highest Pearson index and the one with the highest
Spearman's index (both in terms of absolute value).\par
Failing to reach the value $0.2$ for neither index,
the results are omitted because irrelevant. However, this failed attempt
motivates the choice to continue the investigation focusing on the only
atmospheric pressure values, since neither temperature nor speed
of the wind seem to have a particular correlation with the number of
hospitalizations.

\section{Use of half-hourly pressure values}
In this study, the pressure measures show that small, short and continuous pressure variations are present.

In order to study the effects of these variations could have on the incidence of pulmonary embolism cases, it was seen that was no longer sufficient to analyse the daily averages of atmospheric pressure. 

For this reason an analysis carried out using pressure values recorded every half an hour. (i.e. 48 daily records, during the period 2016-2018, data provided by the Lamma Consortium).\\

\subsection{Within a day variations}
\label{section:sbalzi}
In accordance with the hypothesis that pressure changes are determinants for human health, the total pressure variation during the day was calculated for each day belonging the sample interval. 

This methodology was found to be more effective than analyzing peaks and patterns. 

The choice of this methodology was also due to empirical observations. 

Imagine having a glass tube full of fluid and an impurity not occluding the lumen of the vessel, for example: a gas bubble.

To detach the impurity from the wall and then remove it, is more effective to proceed applying a series of repeated and delicate taps to the tube, rather than a single strong shot.

By analogy, it was thought that, a series of repeated, small and rapid changes in pressure were more probable causes of a thrombus detachment from the venous wall, rather than a single surge of strong entity.\par
The calculation of the variations has been done in the following way: for every day $d$, given the half-hourly values $d_1, d_2, ..., d_{48}$,the daily pressure change $\Delta_d$ has been calculated as:
	$$\Delta_d = \sum_{k=2}^{48} |d_{k} - d_{k-1}|$$\par
A graph has then been drawn showing the averages of the variations $\Delta_d$ depending on the number of daily hospitalizations, on the same day and in the three preceding days. 

Initially it was considered a test data set only the period corresponding to the year 2016.

\begin{figure}
	\makebox[\textwidth][c]{\includegraphics[width=\textwidth]{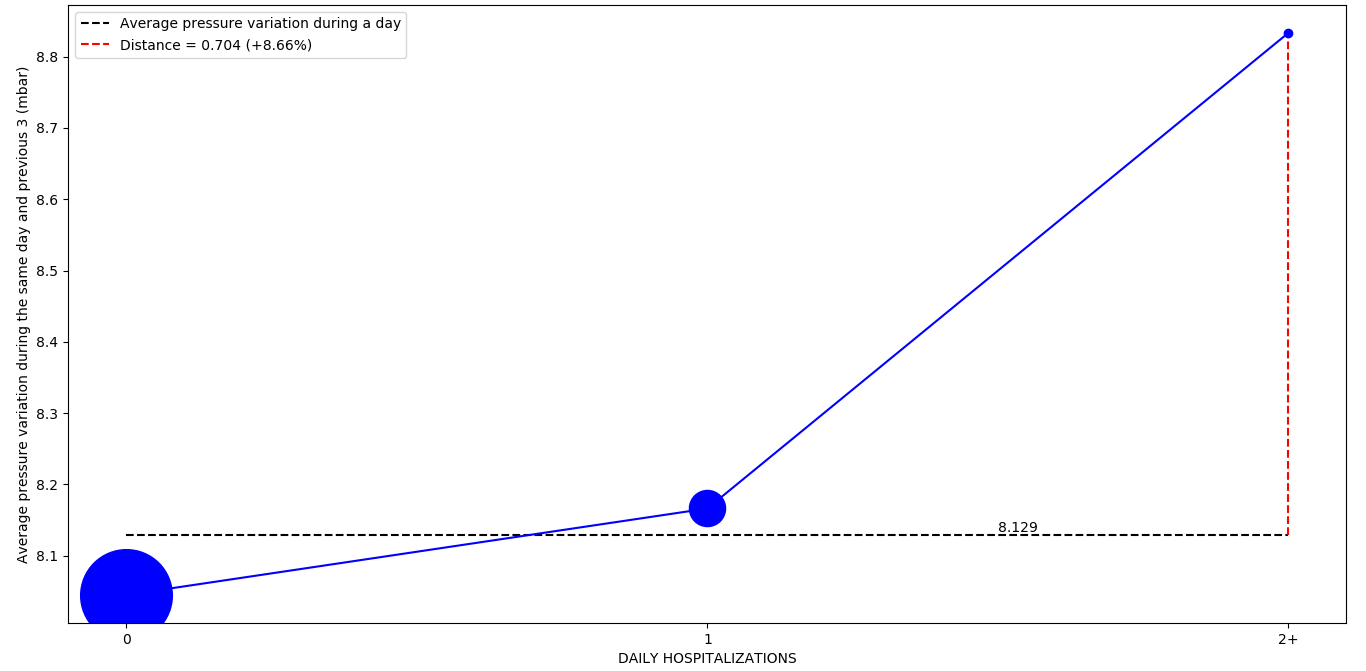}}
	\caption{Average pressure changes in 4 days with half-hourly records divided by number of hospitalizations at 4 th day ( test period 2016 )}
	%\captionsetup{justification=centering,margin=2cm}
	\centering
	\label{fig:pres_var_test}
\end{figure}

  The result is in Figure \ref{fig:pres_var_test} (the size of the points in the graph is directly proportional to the number of occurrences).
  
  This graph shows that, in the periods immediately preceding the days with two or more hospitalizations there is an average variation of pressure $\Delta_d$ higher of 8.66\% than to the average variation over the whole period considered.
    
This result seems to confirm the hypothesis that a greater number of pressure changes correspond, in the short term, an increase in cases of pulmonary embolism.\par
For a better confirmation, the procedure has been repeated on all the data in our possession, that is on all the 2016-2018 period.

The analysis we have done, seem to confirm only partially what was found.

\begin{figure}
	\makebox[\textwidth][c]{\includegraphics[width=\textwidth]{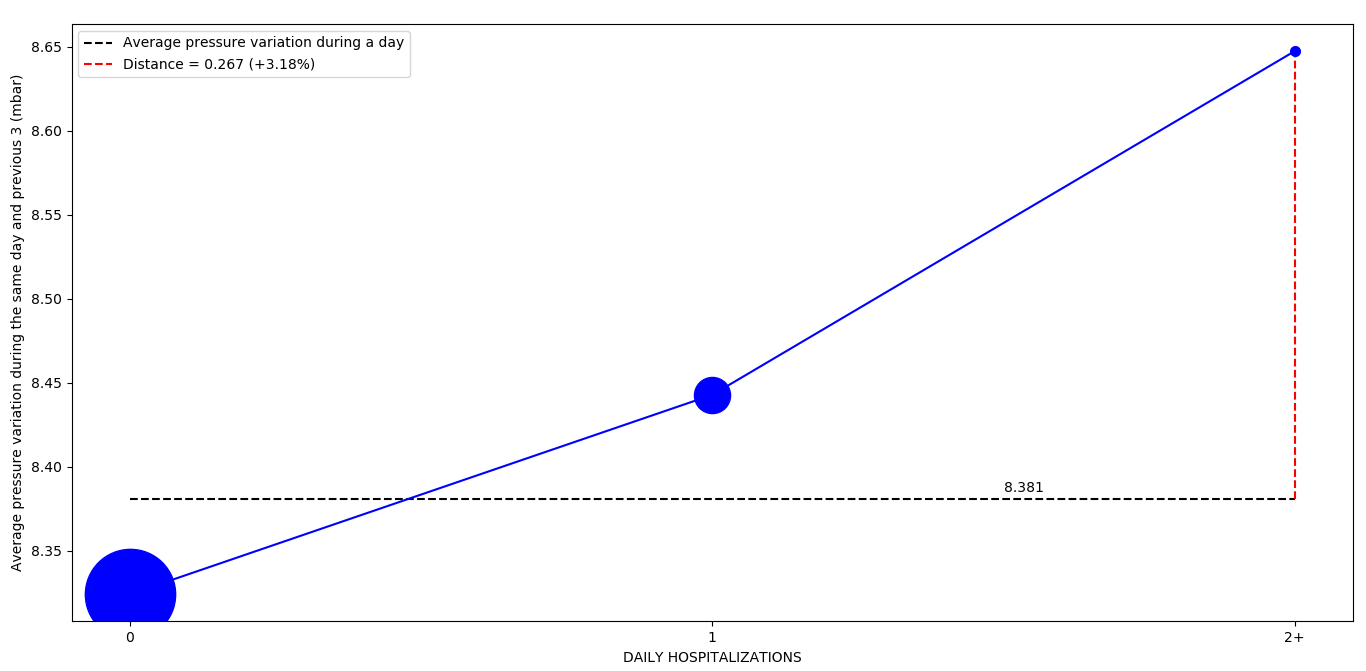}}
	\caption{Average pressure changes in 4 days with half-hourly records divided by hospitalization numbers on the 4th day (2016-2018 period)}
	%\captionsetup{justification=centering,margin=2cm}
	\centering
	\label{fig:pres_var}
\end{figure}

The graph shown in Figure \ref{fig:pres_var} is more flattened, and shows an increase in the variation in the periods before the days with more hospitalizations equal to only 3.18\%.

\par
All this is implemented in the script \cite[\texttt{daily\_variation.py}]{github:weape}.

\subsubsection{ Results at different daily intervals}
Only the graphs obtained by analyzing the averages on the time interval that produced the best results over the whole period 2016-2018, i.e. four days, have been reported.
For completeness, the results obtained for some of the other ranges analysed are given in Table \ref{tab:percent}.

In particular, it shows the percentage increase of the average pressure changes in the periods before the days with two or more hospitalizations compared to the average over the whole period considered.

Although with more or less incisive results, it should be noted that there is an increase independent from the amplitude of the range and the length of the sample analysed.

Finally, it can be observed that, by extending the length of the sample, the increase is always scaled down to three years.

\begin{table}
\begin{tabularx}{\textwidth}{XXX}
%\toprule
\hline
Length of interval in days & Percentage increase in the year 2016 & Percentage increase in the years 2016-2018\\
\hline
%\midrule
2 & 9.77\% & 0.16\% \\
3 & 5.64\% & 2.13\% \\
\rowcolor{gray!20}
4 & 8.66\% & \textbf{3.18\%} \\
5 & \textbf{10.61\%} &  3.08\% \\
6 & 9.24\% & 2.78\% \\
7 & 7.43\% & 1.86\% \\
%\bottomrule
\hline
\end{tabularx}
\caption{Percentage increase in pressure changes in the preceding periods - days with two or more hospitalizations in relation with the average of the variations - over the whole period.}
\label{tab:percent}
\end{table}

\subsection{Study of spectrograms}

Having increased, at least for the pressure, the sample detection frequency, we now have a much larger data set.

This fact allowed to try to apply the Fourier Fast Transform to calculate the spectrograms of the historical series of pressure and that of the hospitalizations \cite{scipy:fft}.

In order to have two series of the equal length and to compare more easily the spectrogram, it was thought to expand also the series of hospitalizations, using a formal artifice, to semihourly samples.

Since the information on the distribution of hospitalizations within the day was unknown, it was chosen to assume a homogeneous distribution.

For this reason, each daily observation has been divided into $48$ semi-hourly observations, each of $1/48$ of the daily hospitalizations.

\begin{figure}
	\makebox[\textwidth][c]{\includegraphics[width=\textwidth]{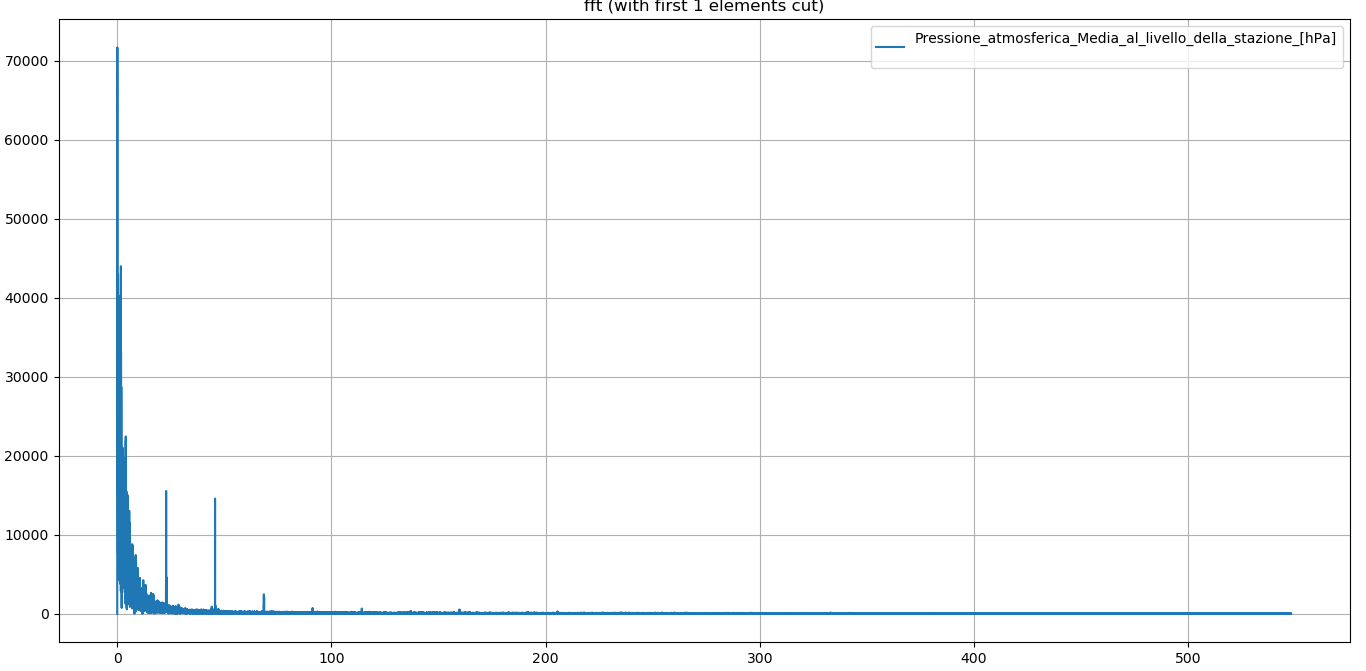}}
	\caption{Spectrogram of atmospheric pressure}
	%\captionsetup{justification=centering,margin=2cm}
	\centering
	\label{fig:fft_pres}
\end{figure}

\begin{figure}
	\makebox[\textwidth][c]{\includegraphics[width=\textwidth]{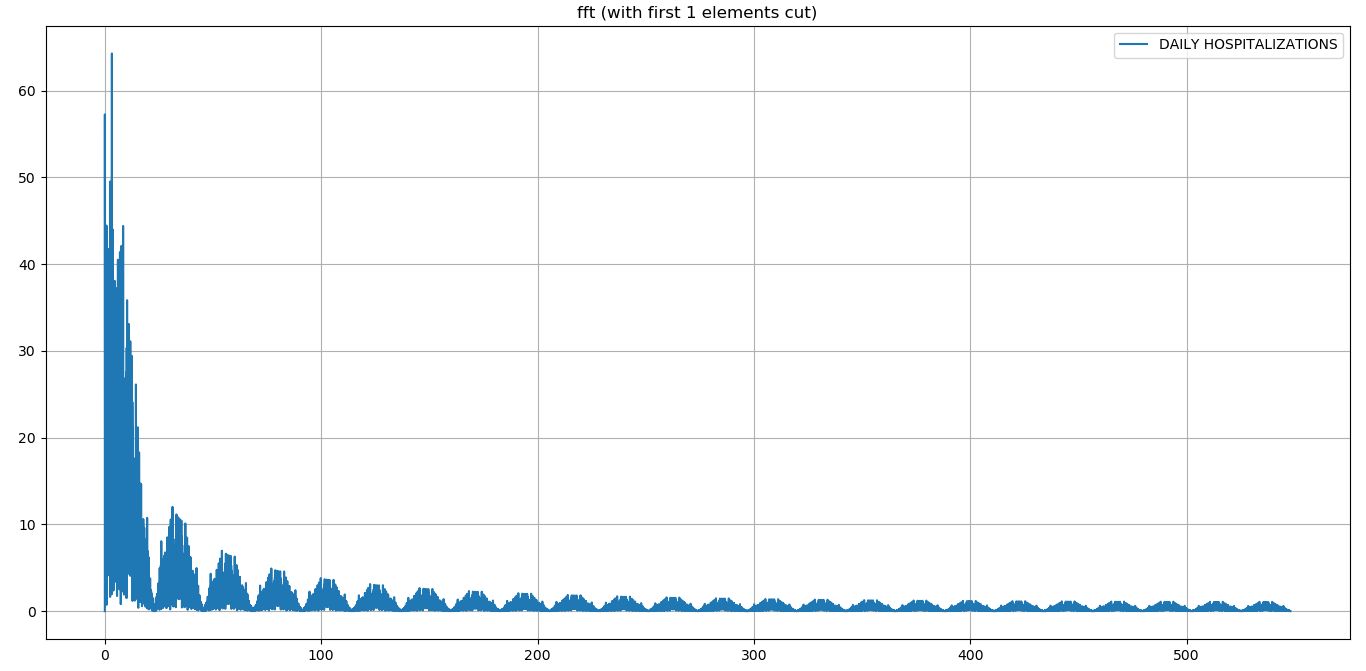}}
	\caption{Spectrogram of hospitalizations}
	%\captionsetup{justification=centering,margin=2cm}
	\centering
	\label{fig:fft_hosp}
\end{figure}

The two spectrograms obtained are those in Figures \ref{fig:fft_pres} and \ref{fig:fft_hosp}.

In these figures the frequency on the $x$-axis is expressed in $1/days$.

Note that, to make the graphs more readable, in both figures the value of the first frequency $(0hz)$ has been manually set to $0$.

This was done because, for the properties of the treated signals, it would normally have a very high value.

To see the algorithm used in detail and the graphs with the first frequency values left unchanged, see paragraph \cite[\texttt{fft.py}]{github:weape}.

No relationship or analogy seems to emerge by comparing the two historical series in the domain of frequencies. 

This could also be due to the method by which the series of hospitalizations was expanded.

Therefore, no particular conclusion can be drawn with a sample of this size.

\section{Conclusions}
Many of the attempts we have done, lead to no statistically significant results, thus confirming the difficulty of the existing literature to give a definite answer on the subject.

Despite this, the research carried out highlighted the strong correlation between the moving averages of atmospheric pressure and those of the number of hospitalizations discussed in Section \ref{section:correlazione}.

The existence of this correlation, from the results obtained, is undeniable for the sample studied.

As already mentioned, however, a possible causality between the time series of pressures and that of hospitalizations is far from certain.

This is due to the fact that the possible factors that could have caused the increase in the number of hospitalizations are manifold, and many of these, probably, do not concern meteorological factors, such as, for example, the improvement of diagnostic tools.

For this reason, the result obtained is a solid result, but it is necessary to start more studies in this field, which analyze different samples.

It would be interesting to see if, in a geographical area ( both the same and other areas ) which has, during a period of time similar to the one studied ( 3 years ), an increase in the annual moving average of pressure atmospheric, a decrease in the moving average number of hospitalizations occurs.

In this case, many of the causes of other nature, thus obtaining a more certain answer.

Such efforts should be a priority in any future developments given the severity of the disease and the difficulty of its diagnosis.

Another significant finding is: the existence of an increase in the number of hospitalizations in the days following short-to-medium periods of time characterized by a high number of half-hourly pressure changes, observed in Section \ref{section:sbalzi}.

Results obtained seems to give credit to the hypothesis of considering the physical phenomenon of thrombus detachment as the effect of very small pressure variations recurring.

However, unlike the moving average, this is not an unequivocal result.

This considering that the result over the whole period 2016-2018 is much more contained than the one related to 2016 only.

In this sense it would be interesting to confirm or deny our results by studying what happens for longer periods of time, covering several years.

The study of the spectrograms obtained by the Fourier transformation will undoubtedly be at the centre of future developments: data in our possession (especially hospitalization data) were found to be too few to carry out analyses of this type, but better results could emerge by repeating the procedure on larger datasets.

In conclusion, although further confirmations are needed, there seems to be some kind of correlation between pulmonary embolism and meteorological parameters, and in particular,  atmospheric pressure seems to be more relevant than temperature and wind speed, which, moreover, is strongly related to pressure variations.
\newpage

%%%%%%%%%%%%%%%%%%%%
%%%% APPENDICI %%%%%%%%%%
%%%%%%%%%%%%%%%%%%%%

\section{Appendix A: WeatherPE}
	
\subsection{Purpose of the appendix}
This appendix wants to be a tool to help you understand some scripts and project functions in Python \emph{Weatherpe}.

This project was created to carry out the analysis of this study, in order to be modular and applicable to the analysis of other statistical samples, which were mentioned in the document.

This appendix is not intended as documentation or a full explanation of the code, for that see the code itself \cite{github:weape}.

\subsection{Structure of the project}
The project is divided into four folders:
\begin{itemize}
	\item \texttt{res}: Contains the data files used;
	\item \texttt{weape}: Contains the classes and functions used;
	\item \texttt{launchers}: Contains executable codes to generate all results included in this document;
	\item \texttt{tests}:Contains tests of methods and functions.
\end{itemize}

\subsection{Series}
File \texttt{series.py} implements \texttt{Series} class, which has been used to represent historical series. This file includes only two attributes:: \texttt{values} and \texttt{label} storing respectively a list containing the values of the series and a string containing the name of the same.\\
Below you will be documented the most important methods at the end of the understanding of the document.

\subsubsection{Normalise}
\label{section:normalise}
Due to the various meanings it assumes in different scientific disciplines, the concept of normalisation always causes some ambiguity, unless it is explicitly stated\cite{treccani:enciclopedia}. For this project, the term normalization has been used to mean a transformation of random variables, also known as min-max normalization, described by the following formula:
\begin{equation}
	z_i = \frac{x_i - min}{max - min}(b-a) + a
\end{equation}
where $x_i$  is a generic value of the random variable $X$, $min$ and $max$ are respectively the minimum and maximum values assumed by $X$ and $[a,b]$ is the new range within which the values of the random variable will be scaled.
this transformation is implemented in an efficient and intuitive way in the normal method \texttt{normalise(self, feature\_range)}, where the \texttt{feature\_range} pair, which if not specified is equal to $(0,1)$, indicates the exremes $a$ and $b$.

\subsubsection{Variation\_series}
\label{section:variationseries}
\texttt{variation\_series(self, length: int, unsigned: boolean)}, 
executed on an object of type \texttt{series} representing a historical series of length $n$, returns an object of the same type representing a historical series of length  $n-length+1$ where the value at instant $t$ contains:

\begin{itemize}
	\item in the case \texttt{unsigned = False} (default option), the difference between the first and the last value of the interval $[t, t+length]$ of the starting time series; 
	\item in the case \texttt{unsigned = True}, the sum of the absolute values of the differences between each value of the starting time series and the next one, within the interval $[t, t+length]$.
\end{itemize}

\subsubsection{Mobile\_mean}
\texttt{mobile\_mean(self, window: int, ws: list)}, executed on an object of type \texttt{series} representing a time series of length $n$, returns an object of the same type representing a time series of length $n-window + 1$, where the value at instant $t$ contains the mobile average weighted according to the weights contained in \texttt{ws} over the interval $[t, t+window]$. If the weights are not specified, the arithmetic moving average is executed.

\subsubsection{Eventuality}
\texttt{eventuality(self)}, executed on an object of type  \texttt{series} representing a time series of length $n$, returns an object of the same type representing a time series of the same length, where the value at instant $t$ contains:

\begin{itemize}
	\item $0$ if the starting time series is also $0$ at instant $t$;
	\item $1$ otherwise.
\end{itemize}

\subsection{PeakSeries}
\texttt{PeakSeries} extends \texttt{Series}, with the additional method \texttt{pattern\_series(self, pattern: tuple)} generating the time series containing the occurrences of a given peacks pattern. Objects of this class are generated by an object of type  \texttt{PeakMaker} by means of the method \texttt{get\_peaks(self)}.

\subsection{Correlation}
It is a class composed of two objects of the \texttt{Series} class that provides different methods to calculate correlation indices between the two time series, to draw graphs of various types, or to translate one series with respect to the other.

%%%%%%%%%%%%%

\section{Appendix B: difinitions of medical and meteorological terms}
\label{section:termini medici}
\subsection{Medical terms}

\subsubsection{Thrombus}
A thrombus, colloquially called blood clot, is a semisolid substance consisting of cells and fibrin that can locate anywhere in the circulatory district, such as arteries and veins and is attached to the inner wall of the blood vessels. A clot is a healthy response to injury to prevent bleeding, but, when clots obstruct blood flow through healthy blood vessels, it can become the leading cause of some severe pathologies in which case we talk about thrombus \cite{zangara:medicina}.

\subsubsection{Deep vein thrombosis}
When thrombosis occurs within deep blood veins, usually at the level of the lower limbs, we speak of deep vein thrombosis (DVP). \cite{msd:tvp}

\subsubsection{Embolism}
The embolism occurs when a thrombus is detached from the wall of a blood vessel to which it is attached. The thrombus or some of its parts can enter into blood circulation until it stops in a blood vessel smaller than the source vessel reducing the blood supply to the downstream tissues \cite{zangara:medicina}.

\subsubsection{Pulmonary embolism}
Pulmonary embolism (PE) is a blockage of an artery in the lungs by a clot that has moved from other districts in the body through the bloodstream (embolism). PE usually results from a blood clot in the leg that travels to the lung. Main signs of the PE include low blood oxygen levels, rapid breathing, rapid heart rate, that cause circulatory and respiratory problems. 
In most cases, the PE is preceded by deep vein thrombosis (DVT). PE and DVT share both the risk factors and the triggers. Among these, advanced age and some pathological conditions have been found to be the main risk factors. In Italy, PE occurs in one patient per 100000 and is the cause of about 15\% of hospital deaths, which rise to 30\% if it is not treated correctly \cite{terzano:malattie}.

\subsubsection{Venous thromboembolisms}
Pulmonary embolism and deep vein thrombosis are two closely related pathological manifestations, which can be described by a single pathological process that is known as venous thromboembolism (VTE) or thromboembolism \cite{treccani:tev}.

\subsection{Meteorological terms}
\subsubsection{Atmospheric pressure}
Atmospheric pressure measures the total weight exerted on a horizontal unit surface by the air column above \cite{giuliacci:manuale}. The atmospheric pressure is measured with an instrument called a barometer. Generally, atmospheric pressure is measured in atmospheres (atm) or millbar (mbar). However, neither of these two units of measurement is the adopted unit by the System, which instead adopts the pascal $(Pa)$ for the measurement of pressure \cite{bipm:si}. In this study the adopted unit of measurement of atmospheric pressure is the millibar. One millibar corresponds to $100 Pa$.

\subsubsection{Average daily atmospheric pressure}
Average daily atmospheric pressure is the arithmetic mean among all atmospheric pressure values recorded over a full day that is normally detected from 00:00 to 23:59. The number of recordings are made at regular intervals throughout the day to achieve a discretization of the number of registrations.

\subsubsection{Variation of atmospheric pressure regimes}
Although atmospheric pressure values in a given area tend to remain the same over the long term, these pressure values can change from day to day or from month to month due to weather phenomena. Different average values of annual pressures are also possible due to climate factors.

\end{document}